\documentclass[twocolumn]{aastex62}

\usepackage{booktabs}

\submitjournal{ApJ}

\title[ConvNets and DES lenses]{An extended catalog of galaxy-galaxy strong gravitational lenses
discovered in DES using convolutional neural networks}
\shortauthors{Jacobs et al}

\begin{document}

\title{An extended catalog of galaxy-galaxy strong gravitational lenses
discovered in DES using convolutional neural networks}

\author{C. Jacobs}
\affiliation{Centre for Astrophysics and Supercomputing, Swinburne University of
Technology, P.O. Box 218, Hawthorn, VIC 3122, Australia}
\affiliation{ARC Centre of Excellence for All Sky Astrophysics in 3 Dimensions (ASTRO 3D),
Swinburne University of Technology, Hawthorn, VIC 3122, Australia}
\author{T. Collett}
\affiliation{Institute of Cosmology \& Gravitation, University of Portsmouth,
Portsmouth, po1 3fx, UK}
\author{K. Glazebrook}
\affiliation{Centre for Astrophysics and Supercomputing, Swinburne University of
Technology, P.O. Box 218, Hawthorn, VIC 3122, Australia}
\affiliation{ARC Centre of Excellence for All Sky Astrophysics in 3 Dimensions (ASTRO 3D),
Swinburne University of Technology, Hawthorn, VIC 3122, Australia}
\author{E.~Buckley-Geer}
\affiliation{Fermi National Accelerator Laboratory, P. O. Box 500, Batavia, IL 60510, USA }
\author{H. T.~Diehl}
\affiliation{Fermi National Accelerator Laboratory, P. O. Box 500, Batavia, IL 60510, USA }
\author{H.~Lin}
\affiliation{Fermi National Accelerator Laboratory, P. O. Box 500, Batavia, IL 60510, USA }
\author{C. McCarthy}
\affiliation{School of Software and Electrical Engineering, Swinburne University of
Technology, P.O. Box 218, Hawthorn, VIC 3122, Australia}
\author{A. K. Qin}
\affiliation{School of Software and Electrical Engineering, Swinburne University of
Technology, P.O. Box 218, Hawthorn, VIC 3122, Australia}
\author{C. Odden}
\affiliation{Fermi National Accelerator Laboratory, P. O. Box 500, Batavia, IL 60510, USA }
\affiliation{Phillips Academy, Andover, MA 01810}
\author{M. Caso Escudero}
\affiliation{Fermi National Accelerator Laboratory, P. O. Box 500, Batavia, IL 60510, USA }
\affiliation{Phillips Academy, Andover, MA 01810}
\author{P. Dial}
\affiliation{Fermi National Accelerator Laboratory, P. O. Box 500, Batavia, IL 60510, USA }
\affiliation{Phillips Academy, Andover, MA 01810}
\author{V. J. Yung}
\affiliation{Fermi National Accelerator Laboratory, P. O. Box 500, Batavia, IL 60510, USA }
\affiliation{Phillips Academy, Andover, MA 01810}
\author{S. Gaitsch}
\affiliation{Fermi National Accelerator Laboratory, P. O. Box 500, Batavia, IL 60510, USA }
\author{A. Pellico}
\affiliation{Fermi National Accelerator Laboratory, P. O. Box 500, Batavia, IL 60510, USA }
\author{K. A. Lindgren}
\affiliation{Fermi National Accelerator Laboratory, P. O. Box 500, Batavia, IL 60510, USA }
\author{T.~M.~C.~Abbott}
\affiliation{Cerro Tololo Inter-American Observatory, National Optical Astronomy Observatory, Casilla 603, La Serena, Chile}
\author{J.~Annis}
\affiliation{Fermi National Accelerator Laboratory, P. O. Box 500, Batavia, IL 60510, USA}
\author{S.~Avila}
\affiliation{Instituto de Fisica Teorica UAM/CSIC, Universidad Autonoma de Madrid, 28049 Madrid, Spain}
\author{D.~Brooks}
\affiliation{Department of Physics \& Astronomy, University College London, Gower Street, London, WC1E 6BT, UK}
\author{D.~L.~Burke}
\affiliation{Kavli Institute for Particle Astrophysics \& Cosmology, P. O. Box 2450, Stanford University, Stanford, CA 94305, USA}
\affiliation{SLAC National Accelerator Laboratory, Menlo Park, CA 94025, USA}
\author{A.~Carnero~Rosell}
\affiliation{Centro de Investigaciones Energ\'eticas, Medioambientales y Tecnol\'ogicas (CIEMAT), Madrid, Spain}
\affiliation{Laborat\'orio Interinstitucional de e-Astronomia - LIneA, Rua Gal. Jos\'e Cristino 77, Rio de Janeiro, RJ - 20921-400, Brazil}
\author{M.~Carrasco~Kind}
\affiliation{Department of Astronomy, University of Illinois at Urbana-Champaign, 1002 W. Green Street, Urbana, IL 61801, USA}
\affiliation{National Center for Supercomputing Applications, 1205 West Clark St., Urbana, IL 61801, USA}
\author{J.~Carretero}
\affiliation{Institut de F\'{\i}sica d'Altes Energies (IFAE), The Barcelona Institute of Science and Technology, Campus UAB, 08193 Bellaterra (Barcelona) Spain}
\author{L.~N.~da Costa}
\affiliation{Laborat\'orio Interinstitucional de e-Astronomia - LIneA, Rua Gal. Jos\'e Cristino 77, Rio de Janeiro, RJ - 20921-400, Brazil}
\affiliation{Observat\'orio Nacional, Rua Gal. Jos\'e Cristino 77, Rio de Janeiro, RJ - 20921-400, Brazil}
\author{J.~De~Vicente}
\affiliation{Centro de Investigaciones Energ\'eticas, Medioambientales y Tecnol\'ogicas (CIEMAT), Madrid, Spain}
\author{P.~Fosalba}
\affiliation{Institut d'Estudis Espacials de Catalunya (IEEC), 08034 Barcelona, Spain}
\affiliation{Institute of Space Sciences (ICE, CSIC),  Campus UAB, Carrer de Can Magrans, s/n,  08193 Barcelona, Spain}
\author{J.~Frieman}
\affiliation{Fermi National Accelerator Laboratory, P. O. Box 500, Batavia, IL 60510, USA}
\affiliation{Kavli Institute for Cosmological Physics, University of Chicago, Chicago, IL 60637, USA}
\author{J.~Garc\'ia-Bellido}
\affiliation{Instituto de Fisica Teorica UAM/CSIC, Universidad Autonoma de Madrid, 28049 Madrid, Spain}
\author{E.~Gaztanaga}
\affiliation{Institut d'Estudis Espacials de Catalunya (IEEC), 08034 Barcelona, Spain}
\affiliation{Institute of Space Sciences (ICE, CSIC),  Campus UAB, Carrer de Can Magrans, s/n,  08193 Barcelona, Spain}
\author{D.~A.~Goldstein}
\affiliation{California Institute of Technology, 1200 East California Blvd, MC 249-17, Pasadena, CA 91125, USA}
\author{D.~Gruen}
\affiliation{Department of Physics, Stanford University, 382 Via Pueblo Mall, Stanford, CA 94305, USA}
\affiliation{Kavli Institute for Particle Astrophysics \& Cosmology, P. O. Box 2450, Stanford University, Stanford, CA 94305, USA}
\affiliation{SLAC National Accelerator Laboratory, Menlo Park, CA 94025, USA}
\author{R.~A.~Gruendl}
\affiliation{Department of Astronomy, University of Illinois at Urbana-Champaign, 1002 W. Green Street, Urbana, IL 61801, USA}
\affiliation{National Center for Supercomputing Applications, 1205 West Clark St., Urbana, IL 61801, USA}
\author{J.~Gschwend}
\affiliation{Laborat\'orio Interinstitucional de e-Astronomia - LIneA, Rua Gal. Jos\'e Cristino 77, Rio de Janeiro, RJ - 20921-400, Brazil}
\affiliation{Observat\'orio Nacional, Rua Gal. Jos\'e Cristino 77, Rio de Janeiro, RJ - 20921-400, Brazil}
\author{D.~L.~Hollowood}
\affiliation{Santa Cruz Institute for Particle Physics, Santa Cruz, CA 95064, USA}
\author{K.~Honscheid}
\affiliation{Center for Cosmology and Astro-Particle Physics, The Ohio State University, Columbus, OH 43210, USA}
\affiliation{Department of Physics, The Ohio State University, Columbus, OH 43210, USA}
\author{B.~Hoyle}
\affiliation{Max Planck Institute for Extraterrestrial Physics, Giessenbachstrasse, 85748 Garching, Germany}
\affiliation{Universit\"ats-Sternwarte, Fakult\"at f\"ur Physik, Ludwig-Maximilians Universit\"at M\"unchen, Scheinerstr. 1, 81679 M\"unchen, Germany}
\author{D.~J.~James}
\affiliation{Harvard-Smithsonian Center for Astrophysics, Cambridge, MA 02138, USA}
\author{E.~Krause}
\affiliation{Department of Astronomy/Steward Observatory, University of Arizona, 933 North Cherry Avenue, Tucson, AZ 85721-0065, USA}
\author{N.~Kuropatkin}
\affiliation{Fermi National Accelerator Laboratory, P. O. Box 500, Batavia, IL 60510, USA}
\author{O.~Lahav}
\affiliation{Department of Physics \& Astronomy, University College London, Gower Street, London, WC1E 6BT, UK}
\author{M.~Lima}
\affiliation{Departamento de F\'isica Matem\'atica, Instituto de F\'isica, Universidade de S\~ao Paulo, CP 66318, S\~ao Paulo, SP, 05314-970, Brazil}
\affiliation{Laborat\'orio Interinstitucional de e-Astronomia - LIneA, Rua Gal. Jos\'e Cristino 77, Rio de Janeiro, RJ - 20921-400, Brazil}
\author{M.~A.~G.~Maia}
\affiliation{Laborat\'orio Interinstitucional de e-Astronomia - LIneA, Rua Gal. Jos\'e Cristino 77, Rio de Janeiro, RJ - 20921-400, Brazil}
\affiliation{Observat\'orio Nacional, Rua Gal. Jos\'e Cristino 77, Rio de Janeiro, RJ - 20921-400, Brazil}
\author{J.~L.~Marshall}
\affiliation{George P. and Cynthia Woods Mitchell Institute for Fundamental Physics and Astronomy, and Department of Physics and Astronomy, Texas A\&M University, College Station, TX 77843,  USA}
\author{R.~Miquel}
\affiliation{Instituci\'o Catalana de Recerca i Estudis Avan\c{c}ats, E-08010 Barcelona, Spain}
\affiliation{Institut de F\'{\i}sica d'Altes Energies (IFAE), The Barcelona Institute of Science and Technology, Campus UAB, 08193 Bellaterra (Barcelona) Spain}
\author{A.~A.~Plazas}
\affiliation{Department of Astrophysical Sciences, Princeton University, Peyton Hall, Princeton, NJ 08544, USA}
\author{A.~Roodman}
\affiliation{Kavli Institute for Particle Astrophysics \& Cosmology, P. O. Box 2450, Stanford University, Stanford, CA 94305, USA}
\affiliation{SLAC National Accelerator Laboratory, Menlo Park, CA 94025, USA}
\author{E.~Sanchez}
\affiliation{Centro de Investigaciones Energ\'eticas, Medioambientales y Tecnol\'ogicas (CIEMAT), Madrid, Spain}
\author{V.~Scarpine}
\affiliation{Fermi National Accelerator Laboratory, P. O. Box 500, Batavia, IL 60510, USA}
\author{S.~Serrano}
\affiliation{Institut d'Estudis Espacials de Catalunya (IEEC), 08034 Barcelona, Spain}
\affiliation{Institute of Space Sciences (ICE, CSIC),  Campus UAB, Carrer de Can Magrans, s/n,  08193 Barcelona, Spain}
\author{I.~Sevilla-Noarbe}
\affiliation{Centro de Investigaciones Energ\'eticas, Medioambientales y Tecnol\'ogicas (CIEMAT), Madrid, Spain}
\author{M.~Smith}
\affiliation{School of Physics and Astronomy, University of Southampton,  Southampton, SO17 1BJ, UK}
\author{F.~Sobreira}
\affiliation{Instituto de F\'isica Gleb Wataghin, Universidade Estadual de Campinas, 13083-859, Campinas, SP, Brazil}
\affiliation{Laborat\'orio Interinstitucional de e-Astronomia - LIneA, Rua Gal. Jos\'e Cristino 77, Rio de Janeiro, RJ - 20921-400, Brazil}
\author{E.~Suchyta}
\affiliation{Computer Science and Mathematics Division, Oak Ridge National Laboratory, Oak Ridge, TN 37831}
\author{M.~E.~C.~Swanson}
\affiliation{National Center for Supercomputing Applications, 1205 West Clark St., Urbana, IL 61801, USA}
\author{G.~Tarle}
\affiliation{Department of Physics, University of Michigan, Ann Arbor, MI 48109, USA}
\author{V.~Vikram}
\affiliation{Argonne National Laboratory, 9700 South Cass Avenue, Lemont, IL 60439, USA}
\author{A.~R.~Walker}
\affiliation{Cerro Tololo Inter-American Observatory, National Optical Astronomy Observatory, Casilla 603, La Serena, Chile}
\author{Y.~Zhang}
\affiliation{Fermi National Accelerator Laboratory, P. O. Box 500, Batavia, IL 60510, USA}
\collaboration{(DES Collaboration)}








\begin{abstract}
    We search Dark Energy Survey (DES) Year 3 imaging for galaxy-galaxy
strong gravitational lenses using convolutional neural networks,
extending previous work with new training sets and covering a wider 
range of redshifts and colors.
We train two neural networks using images of simulated lenses, then use
them to score postage stamp images of 7.9 million sources from the Dark Energy
Survey chosen to have plausible lens colors based on simulations. 
We examine 1175 of the highest-scored candidates and identify 152 
probable or definite lenses. 
Examining an additional 20,000 images 
with lower scores, we identify a further 247 probable or definite candidates.
After including 86 candidates 
discovered in earlier searches using neural networks and 26 candidates discovered
through visual inspection of blue-near-red objects in the DES catalog, 
we present a catalog of 511 lens candidates.
\end{abstract}

\reportnum{DES-2018-0435}
\reportnum{FERMILAB-PUB-19-133-AE}





\section{Introduction}\label{sec:introduction}

Gravitational lensing is a phenomenon arising from the relativistic
curvature of spacetime around massive objects \citep[see
\citet{treu_strong_2010} for an
overview]{EinsteinLenslikeactionstar1936, ZwickyNebulaeGravitationalLenses1937}.
When strong gravitational lensing occurs, we sometimes observe multiple magnified
images of distant sources that lie behind the lensing mass. When the
lens is a massive galaxy, group or cluster, strong lensing can be detectable
across cosmological distances. Lensing observables such as the Einstein
radius are sensitive to the mass of the lens as well as to
cosmological parameters, lending strong lensing analysis to many
applications across astrophysics and cosmology. 

One of strong lensing's applications is as a precise probe of lens mass,
dark and baryonic, out to redshift 1 and beyond. Early Type Galaxies
(ETGs) contain much of the local universe's stellar mass
\citep{renziniStellarPopulationDiagnostics2006}, and are the majority of
known galaxy lenses due to their high surface mass densities. By
measuring the evolution of the 
total-mass density slopes of ETGs (i.e. constrain the exponent \(\gamma\), 
where \(\rho(r) \propto r^{\gamma}\)),
we can test the two-phase model of galaxy assembly predicted by theorists.
Simulations predict that at early times, gas-rich assembly from
filaments and gas-rich mergers lead to \textit{in situ} star formation,
concentrating baryons in galaxy centres and steepening the density profile.
At later times, mass assembly is dominated by dry minor mergers, depositing
mass on the outskirts of galaxies and thus increasing size, decreasing \(\gamma\)
\citep{wellonsFormationMassiveCompact2015, bellstedtSLUGGSSurveyComparison2018}.
Observations have so far failed to confirm this prediction, with a weak
steepening over time of \(\gamma\) observed instead
\citep{sonnenfeldSL2SGalaxyscaleLens2013, remusCoEvolutionTotalDensity2017}.
At non-local redshifts, galaxy-scale strong lensing remains the only feasible method for
measuring these density slopes. However, the current lens sample is not
large enough to conclusively resolve this tension between simulations and the existing
observations. More galaxy-scale strong lenses are needed, and at higher redshifts.

The statistics of strong lenses may also prove important in ruling in or out
particular models of dark matter. Strong lensing can produce bright
arcs, and in some cases near-perfect Einstein rings. These rings and
arcs are perturbed if they intersect sub-structure within the lens's
dark matter halo, producing detectable `kinks' in the ring. 
The strong lens system SDSS J120602.09+514229.5 described in
\citet{vegettiQuantifyingDwarfSatellites2010a} contains a visible dwarf
that lies on the Einstein ring and introduces a visible distortion in the ring;
the same effect will be detectable for dark subhaloes.
Exploiting this effect using a large sample of bright arcs and rings will allow 
us to constrain the subhalo mass function 
\citep{koopmansGravitationalImagingCold2005, 
vegettiStatisticsMassSubstructure2009}, a technique demonstrated by 
\citet{vegettiInferenceColdDark2014} on 11 strong lensing systems
resulting in a single detection.
\citet{liConstraintsIdentityDark2016} use simulations to calculate that as 
few as 100 bright arcs, with sufficient image
resolution to detect subhaloes down to \(10^7h^{-1}\textsc{M}_\odot\) (consistent
with future observations) would tightly constrain the subhalo cutoff mass.
Such analysis may confirm the \(\Lambda\)CDM paradigm by providing 
direct evidence for the low-mass subhalos predicted by theory; 
conversely, a detection of low-mass subhaloes could provide 
strong evidence for a warm dark matter candidate such as a keV-mass sterile neutrino.
Despali et al. \citeyearpar{DespaliModellinglineofsightcontribution2018}
have also demonstrated a method to constrain the subhalo mass function
using lensing and line-of-sight substructure.

Double-source plane lenses, where two strongly-lensed sources at
different redshifts are detectable, can function as unique cosmological
probes. The ratio of the Einstein radii of the two lenses,
\(\beta\), is independent of the Hubble parameter but sensitive
to the dark energy equation of state, \(w\) and to both
\(\Omega_M\) and \(\Omega_k\). 
\citet{CollettCosmologicalconstraintsdouble2014} used a model of
double-source plane lens SDSSJ0946+1006 to constrain \(w\) with 30 per cent greater precision
than \emph{Planck} alone. Only a few examples of such lenses have been
discovered so far
\citep{GavazziSloanLensACS2008, 
tanakaSpectroscopicallyConfirmedDouble2016,  
diehlBrightArcsSurvey2017}.

Artificial Neural Networks (ANNs) are the key machine learning technique
that underpins recent advances in so-called `Deep Learning'. An overview
of ANNs and deep learning can be found in
\citet{schmidhuberDeepLearningNeural2015a}. Convolutional Neural
Networks \citep[CNNs;][]{lecunBackpropagationAppliedHandwritten1989} are a
type of ANN optimised for problems involving image data. For standard
computer vision tasks such as image classification and object detection,
CNNs have proven highly successful and now routinely exceed human
performance. CNNs have already found many successful applications in
astronomy, for instance galaxy morphology classification
\citep[][]{dieleman_rotation-invariant_2015, DaiGalaxyMorphologyClassification2018};
star-galaxy separation \citep{kimStargalaxyClassificationUsing2017, cabayolPAUSurveyStargalaxy2019}; or
identifying quasars from spectra
\citep{buscaQuasarNETHumanlevelSpectral2018}.

Here we are concerned
with finding and exploiting strong lenses on the galaxy and
group scales.  Several hundred examples of 
galaxy-galaxy strong lenses are currently
known \footnote{L.A. Moustakas \& J. Brownstein, priv. comm. Database of
  confirmed and probable lenses from all sources, curated by the
  University of Utah. http://admin.masterlens.org}. Simulations, such as
Collett \citeyearpar{collett_population_2015} predict that several
thousand lenses should be discoverable in current-generation surveys
such as the Dark Energy Survey
\citep[DES;][]{darkenergysurveycollaborationDarkEnergySurvey2016},
Kilo Degree Survey \citep[KiDS;][]{dejongFirstSecondData2015}, and
Subaru Hyper-Suprime Cam \citep{miyazakiLargeSampleShearselected2018}. Although in
decades past most strong lenses were discovered serendipitously or
through visual inspection of an entire survey, the scale of modern
surveys means a more targeted approach is required. Previous strategies
have included searching catalogs for multiple blue sources near red ETGs
\citep{diehlBrightArcsSurvey2017}, modeling all sources as strong lenses
and testing for goodness of fit
\citep{marshall_automated_2009, ChanChitahStronggravitationallensHunter2015};
or recruiting citizen scientists to examine images in quantity
\citep{MarshallSPACEWARPSCrowdsourcing2016, MoreSPACEWARPSII2016}.
Recently, many efforts have employed modern computer vision and machine
learning techniques. Neural nets have been shown to be effective at
distinguishing between simulated lenses and non-lenses
\citep{LanusseCMUDeepLensDeep2017, AvestruzAutomatedLensingLearner2017, HezavehFastAutomatedAnalysis2017}.
Applied to imaging surveys, \citet[][hereafter
Paper 1]{JacobsFindingstronglenses2017} used CNNs to recover several
hundred known lenses and 17 new candidates in an hour of inspection
time, and Petrillo et al
\citetext{\citeyear{petrilloLinKSDiscoveringGalaxyscale2019}; \citeyear{PetrilloFindingStrongGravitational2017}}
used neural networks to discover over 300 candidate lenses in KiDS.

In \citealt{jacobsFindingHighredshiftStrong2019} (hereafter Paper 2) we
presented 84 candidate lenses at redshifts 0.8 and above discovered in 
the Dark Energy Survey Year 3 imaging using convolutional neural networks. 

In the present paper we present the results of a wider search of the DES images.
We apply the technique developed in
Paper 1 and Paper 2 to the DES Year 3 coadd 
images\footnote{Now available publicly as Data Release 1 at https://des.ncsa.illinois.edu/releases/dr1/} 
\citep{AbbottDarkEnergySurvey2018, MorgansonDarkEnergySurvey2018}, using
newly trained networks
and searching for lenses from a wider range of redshifts, morphologies and colors. In
section~\ref{sec:method} we outline the method used to train the
networks and score candidates. In section~\ref{sec:results} we present
the results of the search and discuss the likely completeness of the
search. In section~\ref{sec:conclusion} we offer concluding remarks.

\section{Method}\label{sec:method}
The lenses in the catalog presented in this paper were discovered
using the
methodology presented in Paper 1 and Paper 2. Here we summarize the method and
describe refinements made since the earlier searches. The catalog
presented in this work includes candidates discovered in searches using
variations on the color cuts, network architectures and simulation
parameters employed in previous work. The search described in Paper 2 
used simulations to target strong lenses in a constrained redshift range;
here we refine the simulations and expand the search, targeting discoverable
lenses across all redshifts, morphologies and colors, aiming for a larger
and purer candidate set.
We note variations from the earlier search in the text and in
Table~\ref{tbl:candidates} where appropriate. The method described below
describes the parameters of the latest and most comprehensive search.

\subsection{Creating simulated lenses}\label{sec:simulations}

In order to train a convolutional neural network to distinguish between
lenses and non-lenses, we require a training set of labeled examples. To
train a network with tens of millions of trainable weights, of the
size required for reliably processing image data, we require a training
set of order tens or hundreds of thousands of labeled examples
\citep[e.g.][]{krizhevsky_imagenet_2012}. Since this exceeds the number
of known lenses by orders of magnitude, we must instead use simulations
to create training sets.

To generate simulations we use the \textsc{LensPop} code described in
\citet{collett_population_2015}. \textsc{LensPop} generates a population
of synthetic elliptical galaxies as lenses, with singular isothermal
ellipsoid mass profiles, using masses, ellipticities and redshifts drawn
from realistic distributions. The lenses are 
simulated with an elliptical De Vaucouleurs profile, and
lensed sources are modelled as
exponential disks with properties drawn from the COSMOS sample
\citep{ilbertCosmosPhotometricRedshifts2009}. Lens images are created
using the \textsc{GRAVLENS} ray-tracing code
\citep{KeetonComputationalMethodsGravitational2001}, with simulated
seeing and shot noise appropriate for DES imaging (see Paper 2). The
simulations produced by \textsc{LensPop} follow a realistic distribution
of lensing parameters such as Einstein radius and magnification; for the
purposes of training the networks we want clear, bright examples of
strong lensing, so we both discard undetectable lenses and make
simulated sources brighter by one magnitude. The thresholds used for detectability are:
signal-to-noise \(> 20\), magnification \textgreater{} 5 and Einstein radius
\textgreater{} \(2\arcsec\). 
Simulated lens and source images are
combined with random patches of DES imaging to create postage stamps
with realistic sky noise, foreground objects, artifacts and other
contaminants.

In addition to the simulations described above and in Paper 2, we create a
further set of simulated lenses using the redMaGiC catalog
\citep{rozoRedMaGiCSelectingLuminous2016} of luminous red galaxies
(LRGs). For each of the 88,307 galaxies in the catalog, we use the
supplied photometric redshift of the galaxy and a nominal velocity dispersion value,
calculated using the \citet{hydeLuminosityStellarMass2009} fundamental plane of SDSS 
ellipticals to convert between rest frame \(r\)-band absolute magnitude 
and velocity dispersion 
(assuming a \(\Lambda\textrm{CDM cosmology, }h=0.7\)). 
We used the redMaGiC photometric redshifts 
and assumed a 10 Gyr old passive SED to convert the observed 
\(i\)-band magnitude into the rest frame \(r\)-band absolute magnitude.
We then sample 3 simulated sources at different positions in the source plane, 
and produce images via raytracing 
for each of the lensed sources. These are then combined with the actual DES imaging
for the galaxy to create simulated lens images.

\begin{figure}
\centering
\includegraphics[width=0.50000\textwidth]{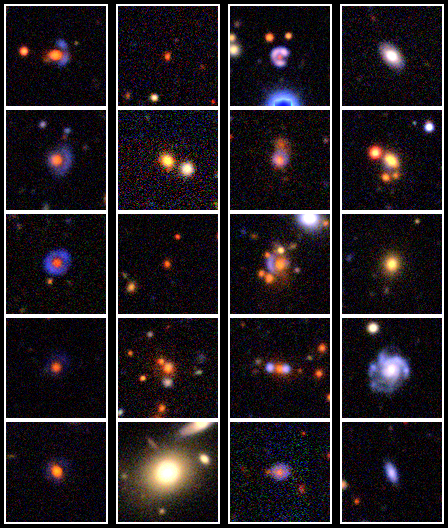}
\caption{Images used in training the neural networks. Left column:
Simulated lenses and lensed sources. Second from left: Simulated ETGs
without a lensed source. Third from left: redMaGiC galaxies and
simulated lensed sources. Right column: Field galaxies, used as negative
examples.}\label{fig:training_images}
\end{figure}

Figure~\ref{fig:training_images} depicts simulations, with and
without synthetic lenses, used for training.

\subsection{Training neural networks}\label{sec:neural_training}

We use the simulations to create training sets for the neural networks.
For the positive examples we use the simulated lenses described in
section~\ref{sec:simulations}. For the negative examples, we use a
combination of simulated non-lenses and real sources from the field. For
the former, we use simulations without any flux from a lensed source
added, depicting only the simulated ETG. For the second, we take postage
stamps of galaxies randomly selected from the target search catalog
(section~\ref{sec:target_catalog}). Since strong lenses are rare - as
few as one in 100,000 galaxies - a random sample of sources from the 
survey catalog is unlikely
to contain any contamination with strong lenses. (We are expecting to 
find a few hundred lenses in our search catalog of 7.9 million galaxies.)

We use two types of negative examples for the following reasons.
Firstly, we use simulated ETGs with and without a lensed source. This way,
the network learns that the presence of a lensed background source is the
significant feature, and that the presence of the elliptical lens
is not in itself indicative of lensing.
Even if the simulated ellipticals are unrealistic in some way, such as 
color, the networks should learn to ignore them as they are present in both
positive and negative examples and are therefore not discriminatory in arriving
at the correct class. The second type of training set, incorporating real 
galaxies as negative examples, exposes the networks to spiral galaxies, mergers,
and other sources that represent non-lenses and will be present in the images
to be tested. In this way the network learns that an elliptical galaxy with
lensing is the target, and that potentially confusing objects such as spiral arms
and tidal tails are to be ignored.

We normalise the training data so that in each
band, the mean value of the supplied image data is zero
and the standard deviation is one; this aids
in quicker convergence of the neural network training process.

We create training sets of up to 200,000 images, consisting of equal
numbers of positive and negative examples. We construct a neural
network with the following architecture: Four convolutional layers, with
kernel sizes 11, 5, 3, and 3 and ReLU\footnote{Rectified Linear Unit: 
\(f(x) = \textrm{max}(0, x)\).}
activations; 2x2 max pooling\footnote{Reduces the spatial extent of the input,
such that for each 2x2 pixel area in the input,
the output is a single value, the maximum of the four values.} after
each convolutional layer; and two fully-connected layers of 1024 neurons
each, with an output layer of two neurons. The network architecture
of CN1 and CN2 is described in detail in Appendix~\ref{sec:appendix-1}.

The process of training, which employs the backpropogation algorithm, is
described in Paper 1, Paper 2 and \citet{lecunBackpropagationAppliedHandwritten1989}. 
Briefly: For each
training example, the algorithm determines a correction to each of the
weights in the network that would decrease a loss function \(L\), where
\(L = 0\) if all classifications are correct and increases as accuracy
decreases. With each iteration, each of the weights of the network are
updated with the mean optimal adjustment calculated
over a batch of 128 training images. By this
process, the network learns key features of the images and converges on
higher classification accuracy. During training we measure the loss and
accuracy on a validation set, consisting of images not shown to the
network during the training steps. This allows us to determine whether the improvements
are a due to an over-fitting to the training set or will generalise well to new
examples.

We train until the validation loss (the loss on the validation set)
does not improve by more than \(10^{-4}\) over six epochs (where
an epoch is a run through the entire training set).

In addition to the networks described in Paper 2, we train four new CNNs
using these training sets, as described in Table~\ref{tbl:searches}.

\hypertarget{tbl:searches}{}
\begin{table*}[ht]
\centering

\caption{\label{tbl:searches}A summary of training sets used to train
neural networks to search DES imaging. redMaGiC sims use real galaxies
for the simulated deflector, LensPop simulates both deflector and lensed
source. Networks TS1 and TS2 are described in Paper 2.}

\begin{tabular}{@{}llll@{}}
\toprule

Network & Positive examples & Negative examples & Training set
size \\\midrule

RM1 & redMaGiC sims & redMaGiC galaxies & 160,000 \\
RM2 & redMaGiC sims & catalog galaxies & 200,000 \\
CN1 & LensPop sims & LensPop sims & 200,000 \\
CN2 & LensPop sims & catalog galaxies & 200,000 \\
TS1 & LensPop high-z sims & LensPop sims & 250,000 \\
TS2 & LensPop high-z sims & Real galaxies & 150,000 \\

\bottomrule
\end{tabular}

\end{table*}

\subsection{Selecting a catalog to search}\label{sec:target_catalog}

\begin{figure}
\centering
\includegraphics[width=0.50000\textwidth]{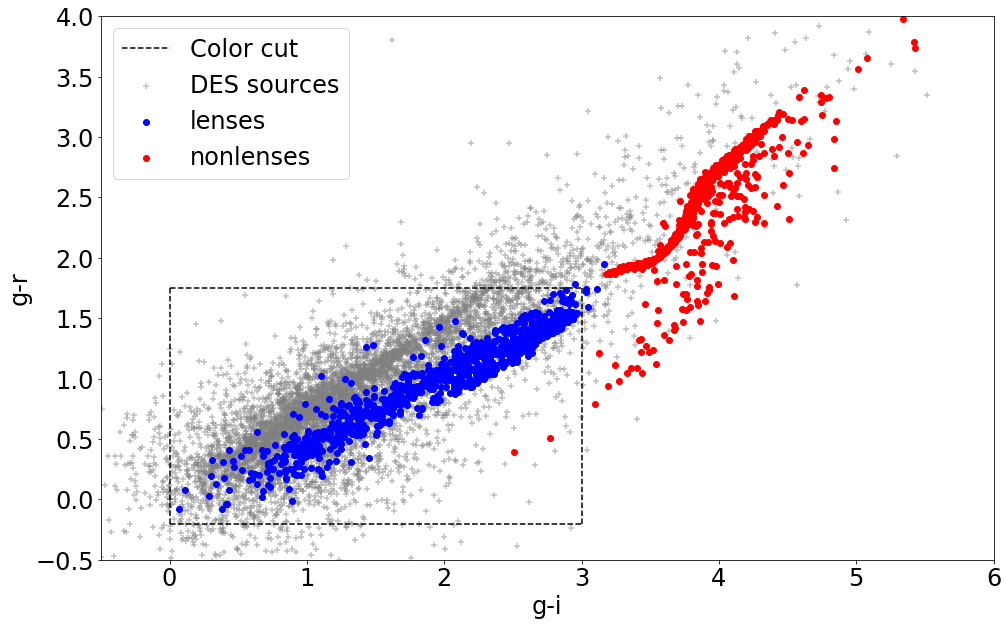}
\caption{Integrated colors of simulated ETGs (blue), simulated strong
lenses (red), and a sample of 5000 sources selected at random from the DES
catalog (grey). The color cuts used to assemble our search catalog are
shown in green. Note that Some simulated galaxies lie at the edge of the DES 
\(g\)-band magnitude limit, resulting in large magnitude errors and
consequent scatter along the diagonal.}\label{fig:sim_colors}

\end{figure}

Even a highly accurate classifier will produce false positives,
especially if it is likely to see irregular objects at classification
time that were not represented in training, which will be the case for
some proportion of sources in any imaging survey. We can minimise false
positives by minimising the number of objects we classify, if we can do
so without discarding any true positives. We restrict our search to only
objects that have the colors of plausible lenses. Although the color of
likely lensing ETGs is well known, the color of combined lens and
lensed source systems in the aperture photometry of the survey catalog
is not known \emph{a priori}. We again turn to simulations as a guide.
Figure~\ref{fig:sim_colors} depicts the combined \(g-r\), \(g-i\)
colors of a sample of 10,000 simulated lenses.

In Paper 2 we searched 1.4 million sources with colors \(2 < g - i < 5\),
\(0.6 < g -r < 3\), or \(1.8 < g - i < 2\), \(0.8 < g -r < 1.2\), based
on simulated lenses at redshifts 0.8 and above. Here we search for
lenses across all redshifts. Using cuts of \(0 < g - i < 3\),
\(-0.2 < g -r < 1.75\), which encompasses 98.7\% of our simulations, we
assemble a catalog of 7.9 million sources from the 300 million objects
in the DES catalog for scoring by the neural networks.

\subsection{Scoring and examining candidates}\label{sec:scoring}

We score 100x100 pixel postage stamp images in \(griz\) bands\footnote{Previous
searches used simulations and survey data in \(gri\) bands only.} with
CNNs trained using the two training sets, RM1 and RM2, described in
section~\ref{sec:simulations}. Each network produces a score in the
interval (0, 1) for each image. We choose thresholds for the CNN scores,
producing lists of candidates with scores greater than those thresholds.
Adjusting this parameter produces candidate sets of varying size, with
different (and unknown) purity and completeness. The thresholds are initially chosen
to produce a candidate set of a few thousand, a convenient size to inspect.
We then determine the purity of the sample, lower the threshold and 
inspect further candidates with lower scores. We repeat this process until
the purity has dropped to the point where diminishing returns (less
than one quality lens candidate per thousand inspections) make
further inspection unworkable.

We inspect RGB images of these
candidates using software, LensRater\footnote{https://github.com/coljac/lensrater}
developed for that purpose, which displays PNG images made with \(gri\)
imaging using three different scaling parameters. We assign each source
a grade from 0-3, where 0 = not a lens, 1 = ``possibly a lens'', 2 =
``probably a lens'', and 3 = ``definitely a lens''. The grades used
throughout the paper represent the mean grade assigned by authors CJ, TC, EBG
and KG.

\subsection{Blue-near-red and rich cluster search}\label{sec:blue-near-red}

In addition to candidates discovered with the CNN search, for 
completeness we include in our catalog 26 
candidates discovered through two visual searches. A search was performed on
53,000 candidates
selected from the DES Y3A1 catalog using a methodology similar to that
described in \citet{diehlBrightArcsSurvey2017}, extended to an extra 3500
square degrees of sky (covering the entire DES footprint in line with the                                           
CNN-based search).   
Then, excluding sources examined in \citet{diehlBrightArcsSurvey2017},
blue-near-red sources were selected from the DES Y3A1 catalog as follows:

\begin{itemize}
    \item Select luminous red galaxies (LRGs) with \(i\)-band magnitude \(< 22\), with redshift \(0.22 < z < 0.70\);
    \item Count blue-colored sources \((-1 \le g-r < 1, -1 \le r-i < 1)\) within \(10\arcsec\);
    \item Examine sources where two blue sources were found near LRGs brighter than 21st magnitude 
    in \(r\), and three or more were found near LRGs brighter than 22nd magnitude.
\end{itemize}

We also examined 759 
sources from the redMaPPer galaxy cluster catalog 
\citep{rykoffRedMaPPerAlgorithmSDSS2014} that matched 
with high-flux sources detected by the Chandra X-ray Observatory
\citep{weisskopfAdvancedXRayAstrophysics1996}.

After visual inspection, and systematic grading according to the prescription above,
40 candidates were graded as ``likely" or ``definite"
lenses, of which 14 were also discovered in the CNN search; 13 are previously known;
five were both rediscovered by the CNN search and previously known; and 26 were new. These candidates are indicated in 
Table~\ref{tbl:candidates}.

\section{Results and discussion}\label{sec:results}

\subsection{Training the network and scoring
sources}\label{sec:training}

\begin{figure}
\centering
\includegraphics[width=0.50000\textwidth]{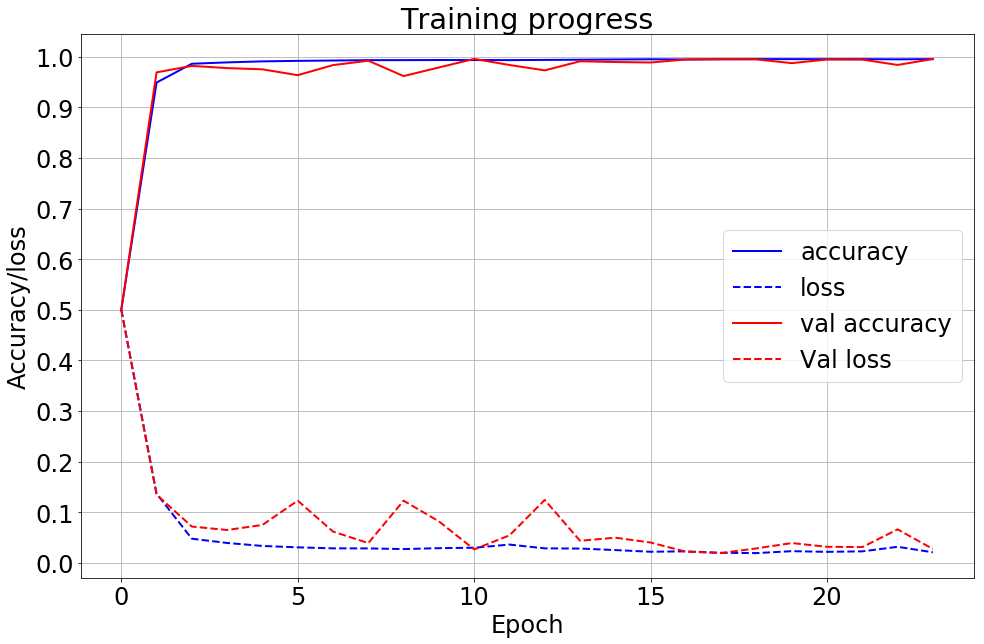}
\caption{Training a neural network on simulated lenses and non-lenses.
Blue dashed line: The loss value, optimized by the training process,
decreasing over time. Red dashed line: The loss evaluated on a
validation set not used for training. Blue solid line: Accuracy in
classification on the training set. Red solid line: Accuracy measured on
the validation set.}\label{fig:training_progress}
\end{figure}

The networks described in section~\ref{sec:neural_training} required
approximately five hours each to train on an NVidia Tesla P90 GPU. 
Training converged on accuracies
of between 99\% and 99.9\% on the validation sets (images not used in the
training process) as shown in Figure~\ref{fig:training_progress}. A typical
Receiver Operating Characteristic (ROC) curve, 
depicting the trade-off between false positives and false negatives on
our training sets, can be found in Paper 2.

Including the overheads of loading images into memory, scoring the 8
million images in our catalog took a total of 20 hours using a Tesla P90
GPU; scoring an individual object required of order \(1 \mu s\). 
scores assigned by the networks are shown in Table~\ref{tbl:scores_distribution}.  

\hypertarget{tbl:scores_distribution}{}
\begin{table}[ht]
\centering

\caption{\label{tbl:scores_distribution}How the convolutional neural
networks scored the catalog of 7.9 million DES sources. The number of
sources scored 1.0 by both RM1 and RM2 was 1175; by CN1 and CN2, 164. }

\begin{tabular}{@{}llll@{}}
\toprule

Network & scores = 0 & scores \textgreater{} .5 & scores = 1 \\\midrule

RM1 & 6248566 & 64525 & 3708 \\
RM2 & 3295227 & 1840383 & 330069 \\
CN1 & 7869097 & 2799 & 1000 \\
CN2 & 2090100 & 1868791 & 228776 \\

\bottomrule
\end{tabular}

\end{table}

\subsection{Selecting candidates}\label{sec:candidates}

We choose candidates to examine by selecting a CNN score threshold and
examining the candidates with a score greater than this number. We seek
a candidate set that is as complete as possible in detectable lenses
included but is of a tractable size and as high a purity (lowest
fraction of false positives) as possible. If the networks are working,
i.e.~the score adds significant information, then we should find fewer
good candidates at lower score values. We first examine a smaller
candidate set, with a high score threshold, and grade the candidates;
then we examine further candidates with lower scores, until diminishing
returns suggest further searching is not feasible. The scores,
subsequent candidate set sizes, and results are summarized in
Table~\ref{tbl:scores_distribution}.

After examining 1175 images with scores of 1.0 by both RM1 and RM2,
we grade 152 with grade \textgreater{}= 2 and a further 148 \textgreater{}= 1. 
We then test for diminishing returns by examining further, lower-scored
sources.  We lower the 
thresholds and examine further candidates.
Using a score threshold of  \textgreater{} 0.5 for both networks, 
we examine a further 15,172 images and grade them as 
247 \textgreater{}= 2, 401 \textgreater{}= 1. 

\begin{figure*}
\centering
\includegraphics[width=0.80000\textwidth]{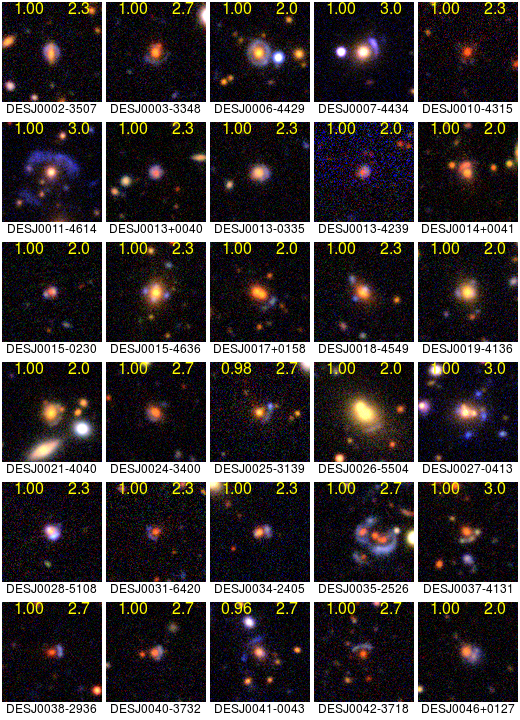}
\caption{Candidate lenses found in DES using
CNNs. In yellow, left: best CNN score, right: human grade.}\label{fig:candidate-lenses-0}

\end{figure*}

Finally, we include candidates from other searches. This includes
networks and catalogs prepared for the search in Paper 2,
and accounts for approximately 20,000 further image inspections. 
We identify a further 86 candidates with grade \textgreater{}= 2 
and 188 with grade \textgreater{}= 1 not identified in the other
inspections.  Including 26 candidates from the rich cluster
and blue-near-red searches, 
we assemble a total catalog of 511 ``probable" and ``definite"
lenses. These candidates are presented in Table~\ref{tbl:candidates}. 
Postage-stamp images of these candidates are presented in Figure~\ref{fig:candidate-lenses-0}, which also depicts 
their CNN scores and human grades. The 742 candidates with grades 
\(>=1\) are presented in Appendix~\ref{sec:appendix-2}, for reference
in future lens searches.


\subsection{Purity and completeness of the candidate catalog}\label{sec:purity}

Previous lens searches have uncovered of order a few hundred potential
lenses in DES. Diehl et al \citeyearpar{diehlBrightArcsSurvey2017}
conducted a search of DES science verification (SV) and Year 1 (Y1)
imaging and identified 374 candidate strong lens systems, approximately
half of which were graded as ``probable" or ``definite" lenses.
The candidates were selected using the
survey catalog, searching for blue-near-red objects and examining a
known catalog of ETGs. Assembling this catalog involved inspection of
approximately 400,000 cutout images. Nord et al
\citeyearpar{nordObservationConfirmationSix2016} searched DES SV and Y1
data for group and cluster-scale strong lenses, identifying 99
candidates of which 21 were confirmed spectroscopically.

On simulated data the CNNs are able to achieve accuracies above 99\% for
a score threshold of 0.5 - in other words, if we consider each candidate
with a score above this value to be a lens. At this rate, we would
expect up to 80,000 false positives from our search catalog in the best
case that the networks were as accurate on real data as on simulations.

Collett's \citeyearpar{collett_population_2015} simulations suggest
\(\sim 1300\) lenses should be findable in DES imaging, using
detectability criteria of a signal-to-noise in \(g > 20\), an Einstein
radius \(> 2\) times seeing, and a magnification of at least 3. However,
blind tests on simulated images of lenses (see Paper 2) indicate that these
detectability criteria may be overly optimistic by a factor of up to
five at estimating what an astronomer can confidently pick from RGB
images. Ideally, we could also examine lens-subtracted images, however
this requires accurate PSF modelling which is complicated for the DES
coadd imaging. Poor PSFs lead to ring-like artifacts in the resultant
imaging, making the discovery of genuine Einstein rings difficult.

\textsc{LensPop} suggests that \(\sim 1300\) lenses should be
detectable; if only one-fifth of these can be confirmed by human
experts, we expect that \(\sim 300\) lenses should be detectable with
confidence. We therefore conclude that our sample is mostly complete for
the survey imaging. As noted below in Section~\ref{sec:comparison}, 41
previously identified high-quality candidates were not recovered
in the CNN search. This indicates that improvements to the networks
would yield further candidates.

If subsequent improvements can be made in
visualisation and grading of DES images, it is possible the detectability
threshold could be lowered and new lenses would become findable. 

\subsection{Suitability of the search catalog}\label{sec:suitability}

\begin{figure}
\centering
\includegraphics[width=0.50000\textwidth]{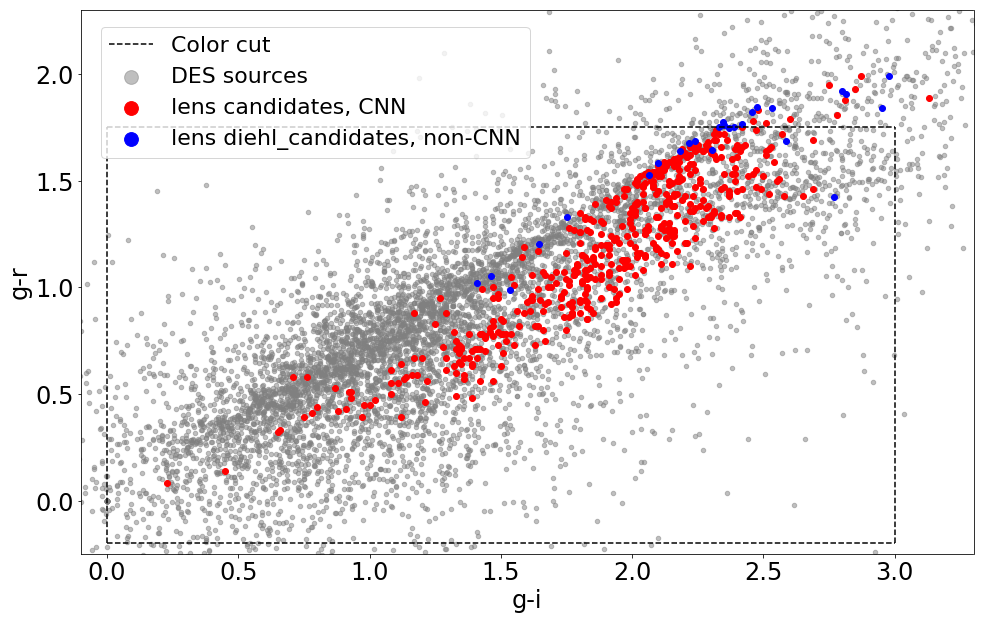}
\caption{Colors of the CNN lens candidates (red), non-CNN lens candidates (blue),
and source catalog (grey).}\label{fig:lens_colors}
\end{figure}

The area of \(g-i\), \(g-r\) color space encompassed by our search
catalog contained 98.5\% of our simulated lenses. 
Figure~\ref{fig:lens_colors} depicts the
location of our candidate lenses in this space. The density of candidates
with \(g-i < 1\) is low, indicating that the search is likely complete
at the blue end of our sample. At the red end, the density of candidates
with \(g-i > 2.5\) also diminishes. Only 27.5\% of our search catalog
lies below this value, but 79\% of our candidates are in this region. 
If we had restricted our search to sources bluer than \(g-i\) of 2.5, 
we would have recovered three-quarters of our best candidates but 
tested only a quarter as many (\(\sim 2\) million) sources. This would
have yielded a purer sample for human inspection, at the cost of some
completeness. Of the 
Diehl et al \citeyearpar{diehlBrightArcsSurvey2017}
candidates, 16 high-quality candidates lay 
outside (redward) of our color cuts by up to
0.25 magnitudes, an indication that future searches would 
benefit from relaxing the criteria.

More candidates lie in the redder end (\(g-r > 1\)), suggesting that
searching further into the red may be worthwhile. However, fewer sources
overall lie in this region of the color space (\(\sim 5\%\) of the DES
catalog are redder than our cuts), and so we expect diminishing returns
to be evident in this area as well.

When we examine the location in this space of candidates with grades of
1 compared to those with a grade of 3, no particular trend is apparent.
More ambiguous candidates in our sample have colors similar to higher-quality
ones.

\subsection{Comparison with blue-near-red and rich cluster search}\label{sec:comparison}

The blue-near-red (BNR) search, using the methodology described in
Section~\ref{sec:blue-near-red} and \citet{diehlBrightArcsSurvey2017},
was able to discover several high-quality lens candidates, but was 
less efficient than the CNN search. Visual inspection of over 
50,000 sources yielded 40 probable or definite lenses, a rate of one 
in 1250; the CNN-based search required the inspection of approximately 
30,000 candidates and yielded 485 probable or definite lenses, a rate 
of one in 62 (as high as one in five in the purest sample).

Of the 26 lenses discovered only in the BNR search, four are of galaxy-galaxy
scale; the remainder are groups and clusters. Since our training set did not
simulate group- and cluster-scale lenses, we do not expect the CNN to
discover these lenses, many of which have Einstein radii larger than the postage 
stamp images scored with the CNNs.

The blue-near-red search methodology was also employed 
in the Diehl et al \citeyearpar{diehlBrightArcsSurvey2017} 
bright arcs survey. 
41 high-quality candidates from that search which satisfied our color cuts
were not recovered in this CNN search
(they received scores below the thresholds we used by 
at least one of our networks). Conversely, of our 
485 CNN-selected candidates, 48  were found by that search.
Identifying those candidates involved visual inspection of over 
400,000 images.

\subsection{False positives}

\begin{figure}
\includegraphics[width=0.50000\textwidth]{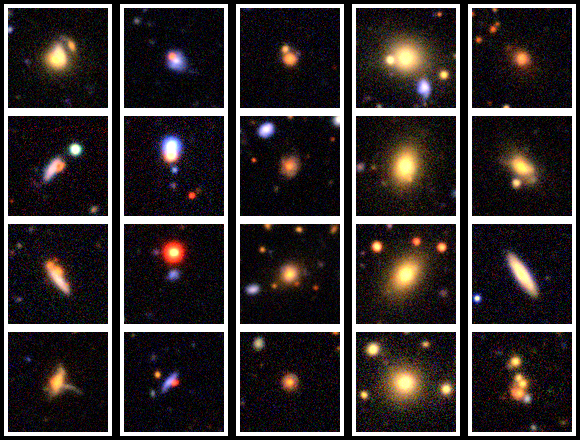}
\caption{False positives scored as definite lenses (score = 1) by a CNN. Left: 
False arcs. Second from left: Blue-near-red objects. Middle: Low signal-to-noise.
Second from right: Bright ETGs. Right: No clear pattern.}\label{fig:false_positives}
\end{figure}

The purest candidate sample inspected yielded 255 quality candidates from 
a sample of 1175, about one in five. Figure~\ref{fig:false_positives} 
depicts some examples of false positives, which fall into four loose 
categories:

\begin{itemize}
    \item False arcs: Arc-like features in the image appear to have
    confused the networks. (5\%)
    \item Blue-near-red: A chance alignment of blue and red sources
    may have confused the network. (24\%)
    \item Low signal-to-noise: Could be a lens but the image is not deep 
    enough to be sure. (10\%)
    \item Unknown: No clear reason (61\%; however, 33\% are bright
    ETGs which could have lensed sources obscured by the lens in the RGB
    imaging inspected by us.)
\end{itemize}

The rate of one-in-five represents a significant improvement from previous searches,
but the fact that a human astronomer would reject the majority of the CNN-selected
candidates clearly implies there is further room for improvement of the method. 

\subsection{Improvements for future searches}\label{sec:improvements}

Our networks produced a candidate set of 1175 candidates with a purity
of \(\sim 13\%\), defined as the proportion of probable or definite
lenses; this figure is greater than 20\% if we include possible lenses. 
Although this represents high accuracy given the number and
variety of sources scored by the networks, with a false positive rate of
1 in 8,000, it suggests that the network could be improved to be more
aggressive in rejecting certain types of candidates that a human would
classify as unlikely. Retraining networks with highly-scored false
positives classified by human inspectors may drop the false positive
rate without significantly impacting the false negative rate. Reducing
the false positive rate would also make wider searches, for instance the
entire survey catalog, more feasible.

The use of transfer learning 
\citep{bengioDeepLearningRepresentations2012, vilaltaTransferLearningAstronomy2018}, 
where a network trained on one particular
problem domain or training set can be applied to a different problem domain
with minimal need for retraining, could assist future searches. The use
of transfer learning for a network with an understanding of galaxy morphology
was demonstrated on SDSS and DES data in \citet{sanchezTransferLearningGalaxy2019}.
Retraining networks trained to find lenses in another survey or with 
larger training sets of known lenses could improve networks used in future searches
of DES or other surveys.

An improvement in the quality of simulations used for training is
also likely to result in improved accuracy. The use of the redMaGiC
simulations resulted in a noticeable improvement in the quality of
candidates, indicating that there was some property of the simulations
that the networks relied on too heavily in scoring. A greater diversity
in synthetic stellar populations, redshifts and morphology may lead to
an improvement in completeness. Our simulations use a PSF drawn from a
distribution consistent with DES Year 1 SV data, which may not be optimal
for the Y3 coadd imaging we searched. This may bias the networks to 
images with seeing closer to the simulated distribution. 
This could also be tested with more varied simulations.

Finally, if the quality of the simulations is not the limiting factor in
performance, then deeper networks may also lead to an improvement. In
theory, larger (more trainable weights) or deeper (more layers) networks
would have the ability to extract more relevant information from the
training sets, but can also prove more difficult to train. Further work
exploring this balance is warranted.

\section{Conclusion}\label{sec:conclusion}

In this paper we present a catalog of 485 strong gravitational lens
candidates discovered in the Dark Energy Survey Y3A1 coadd images using
convolutional neural networks. We used simulated lenses to determine
color cuts, which we applied to the 300 million sources in the DES
survey catalog, yielding 7.9 million sources to search. 
We scored each image with several neural networks trained with
simulated lenses and real galaxies, and combined the scores to produce
small sub-sets of our search catalog that were examined by human inspectors and graded for
quality. Examining one set of 1175 images (0.01\% of the search catalog)
resulted in 152 high-quality (probable or definite lenses). 
Experimenting with networks of different
architectures and training sets, a further approximately 20,000 images
were inspected bringing the total catalog of high-quality candidates to
399. To this we add 86 candidates found in previous CNN searches and
26 new candidates discovered in other visual searches, 
examining rich clusters and blue-near-red sources. 
The 511 candidates we discovered in DES with a grade
\textgreater{}= 2 (``probably a lens'') are presented in Table
\ref{tbl:candidates}. 

For reference by future lens searches, in Appendix~\ref{sec:appendix-2} 
we include 742 lenses with grades \(< 2\), i.e. possible lenses. If 
a significant proportion of these are confirmed in future, 
it may indicate that
the CNNs are able to distinguish lensing features that are difficult 
for a human astronomer to confidently identify.

Future searches will seek to improve the purity of the samples by
retraining networks using discovered lenses and false positives 
graded by human experts including citizen scientist volunteers. 

\section{Acknowledgements}\label{acknowledgements}

This paper has gone through internal review by the DES collaboration.

This research was supported by the Australian Research Council Centre of
Excellence for All Sky Astrophysics in 3 Dimensions (ASTRO 3D), through
project number CE170100013.

TEC is supported by a Dennis Sciama Fellowship from the University of
Portsmouth.

Funding for the DES Projects has been provided by the U.S. Department of Energy, the U.S. National Science Foundation, the Ministry of Science and Education of Spain, 
the Science and Technology Facilities Council of the United Kingdom, the Higher Education Funding Council for England, the National Center for Supercomputing 
Applications at the University of Illinois at Urbana-Champaign, the Kavli Institute of Cosmological Physics at the University of Chicago, 
the Center for Cosmology and Astro-Particle Physics at the Ohio State University,
the Mitchell Institute for Fundamental Physics and Astronomy at Texas A\&M University, Financiadora de Estudos e Projetos, 
Funda{\c c}{\~a}o Carlos Chagas Filho de Amparo à Pesquisa do Estado do Rio de Janeiro, Conselho Nacional de Desenvolvimento Cient{\'i}fico e Tecnol{\'o}gico and 
the Minist{\'e}rio da Ci{\^e}ncia, Tecnologia e Inova{\c c}{\~a}o, the Deutsche Forschungsgemeinschaft and the Collaborating Institutions in the Dark Energy Survey. 

The Collaborating Institutions are Argonne National Laboratory, the University of California at Santa Cruz, the University of Cambridge, Centro de Investigaciones Energ{\'e}ticas, 
Medioambientales y Tecnol{\'o}gicas-Madrid, the University of Chicago, University College London, the DES-Brazil Consortium, the University of Edinburgh, 
the Eidgen{\"o}ssische Technische Hochschule (ETH) Z{\"u}rich, 
Fermi National Accelerator Laboratory, the University of Illinois at Urbana-Champaign, the Institut de Ciències de l'Espai (IEEC/CSIC), 
the Institut de F{\'i}sica d'Altes Energies, Lawrence Berkeley National Laboratory, the Ludwig-Maximilians Universit{\"a}t M{\"u}nchen and the associated Excellence Cluster Universe, 
the University of Michigan, the National Optical Astronomy Observatory, the University of Nottingham, The Ohio State University, the University of Pennsylvania, the University of Portsmouth, 
SLAC National Accelerator Laboratory, Stanford University, the University of Sussex, Texas A\&M University, and the OzDES Membership Consortium.

Based in part on observations at Cerro Tololo Inter-American Observatory, National Optical Astronomy Observatory, which is operated by the Association of 
Universities for Research in Astronomy (AURA) under a cooperative agreement with the National Science Foundation.

The DES data management system is supported by the National Science Foundation under Grant Numbers AST-1138766 and AST-1536171.
The DES participants from Spanish institutions are partially supported by MINECO under grants AYA2015-71825, ESP2015-66861, FPA2015-68048, SEV-2016-0588, SEV-2016-0597, and MDM-2015-0509, 
some of which include ERDF funds from the European Union. IFAE is partially funded by the CERCA program of the Generalitat de Catalunya.
Research leading to these results has received funding from the European Research
Council under the European Union's Seventh Framework Program (FP7/2007-2013) including ERC grant agreements 240672, 291329, and 306478.
We acknowledge support from the Australian Research Council Centre of Excellence for All-sky Astrophysics (CAASTRO), through project number CE110001020, and the Brazilian Instituto Nacional de Ci\^encia
e Tecnologia (INCT) e-Universe (CNPq grant 465376/2014-2).

This manuscript has been authored by Fermi Research Alliance, LLC under Contract No. DE-AC02-07CH11359 with the U.S. Department of Energy, Office of Science, Office of High Energy Physics. The United States Government retains and the publisher, by accepting the article for publication, acknowledges that the United States Government retains a non-exclusive, paid-up, irrevocable, world-wide license to publish or reproduce the published form of this manuscript, or allow others to do so, for United States Government purposes.

This research has made use of the NASA/IPAC Extragalactic Database (NED), which is operated by the Jet Propulsion Laboratory, California Institute of Technology, under contract with the National Aeronautics and Space Administration.

CWJ acknowledges travel support for this work provided by the Astronomical Society of Australia.

\startlongtable
\begin{deluxetable*}{cccccccc}
\tablecaption{\label{tbl:candidates}Strong lens candidates from visual
inspection of the neural network-selected sources, sorted by grade.
Redshifts are BPZ\footnote{http://www.stsci.edu/\textasciitilde{}dcoe/BPZ/} 
photometric redshifts except where indicted with \^{}
where spectroscopic redshifts are substituted. imag is the \(i\)-band magnitude
from the DES Y3A1 catalog aperture photometry. Notes: a) previously
known (74 candidates); b) discovered through blue-near-red and rich cluster search described in 
Section~\ref{sec:blue-near-red}; c) found in both searches; d) found
in previous CNN-based search. {\footnotesize {\sc REFERENCES.} $^1$~\citet{diehlBrightArcsSurvey2017}, $^2$~\citet{bleemSPTClusters2015}, $^3$~\citet{furlanettoSOGRAS2013}, $^4$~\citet{starkCASSOWARY2013}, $^5$~\citet{moreCFHTLSStrongLensing2012},
$^6$~\citet{MoreSPACEWARPSII2016}, $^7$~\citet{cabanacCFHTLSStrongLensing2007}, $^8$~\citet{gavazzi_ringfinder:_2014}, $^9$~\citet{hammerAbell3701991}, $^{10}$~\citet{postmanAbell3832012}, $^{11}$~\citet{baylissArcs2012},
$^{12}$~\citet{linDESJ04082017}, $^{13}$~\citet{baylissSPTClusters2016}, $^{14}$~\citet{menanteauSCSClusters2010}, $^{15}$~\citet{bleemBCSClusters2015}, $^{16}$~\citet{abellCatalog1989}, $^{17}$~\citet{nordObservationConfirmationSix2016}, $^{18}$~\citet{buckleygeerElliotArc2011},
$^{19}$~\citet{sonnenfeldSuGOHI2018},
$^{20}$~\citet{kostrzewaCASSOWARY2014}}}
\tablewidth{700pt}
\tabletypesize{\scriptsize}
\tablehead{
\colhead{Candidate} & 
\colhead{objectid} & 
\colhead{RA} & 
\colhead{dec} & 
\colhead{grade} & 
\colhead{photoz} & 
\colhead{imag} & 
\colhead{notes}
}
\startdata
DESJ0002-3507  &  139741252  & 0.59845 & -35.12122
& 2.3 & 0.51 & 19.2 &   \\
DESJ0003-3348  &  139823797  & 0.81825 & -33.80120
& 2.7 & 0.68 & 19.3 &   \\
DESJ0006-4429  &  142775105  & 1.68592 & -44.49735
& 2.0 & 0.53 & 18.6 & a$^{1}$ \\
DESJ0007-4434  &  142779522  & 1.87201 & -44.57949
& 3.0 & 0.52 & 18.1 & a$^{1}$  \\
DESJ0010-4315  &  182452355  & 2.62678 & -43.25413
& 2.3 & 0.84 & 19.9 &   d \\
DESJ0011-4614  &  182003535  & 2.97135 & -46.23942
& 3.0 & 0.60 & 18.5 & a$^{1,2}$  d \\
DESJ0013+0040  &  179698697  & 3.29016 & 0.66767
& 2.3 & 0.75 & 20.0 &   \\
DESJ0013-0335  &  180799530  & 3.29224 & -3.59597
& 2.3 & 0.53 & 19.1 &   \\
\enddata
\end{deluxetable*}




\bibliographystyle{mnras}
\renewcommand\refname{References}
\bibliography{./catalog.bib}

\begin{figure*}
\centering
\includegraphics[width=0.80000\textwidth]{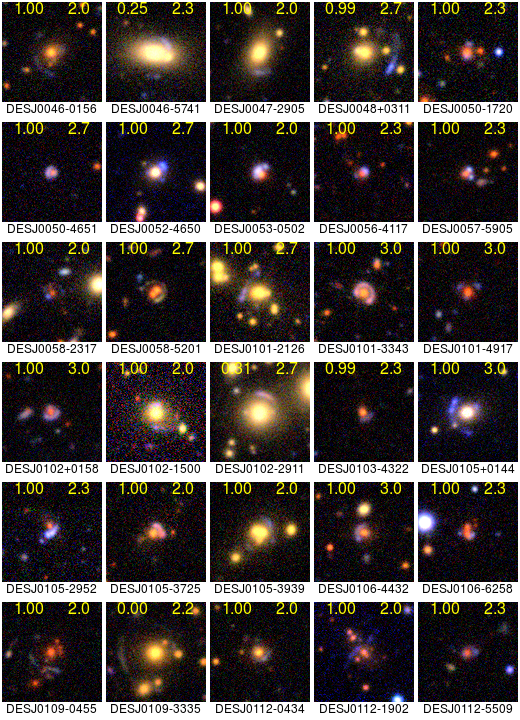}
\caption{Candidate lenses found in DES using
CNNs. In yellow, left: best CNN score, right: human grade.}\label{fig:candidate-lenses-1}

\end{figure*}

\begin{figure*}
\centering
\includegraphics[width=0.80000\textwidth]{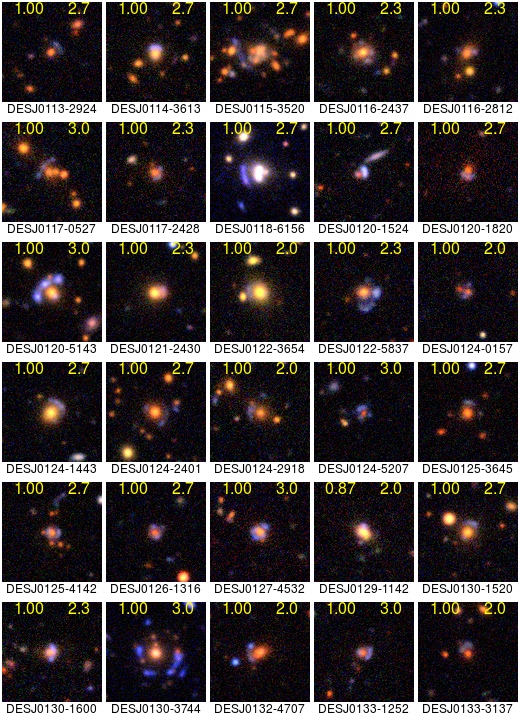}
\caption{Candidate lenses found in DES using
CNNs. In yellow, left: best CNN score, right: human grade.}\label{fig:candidate-lenses-2}

\end{figure*}

\begin{figure*}
\centering
\includegraphics[width=0.80000\textwidth]{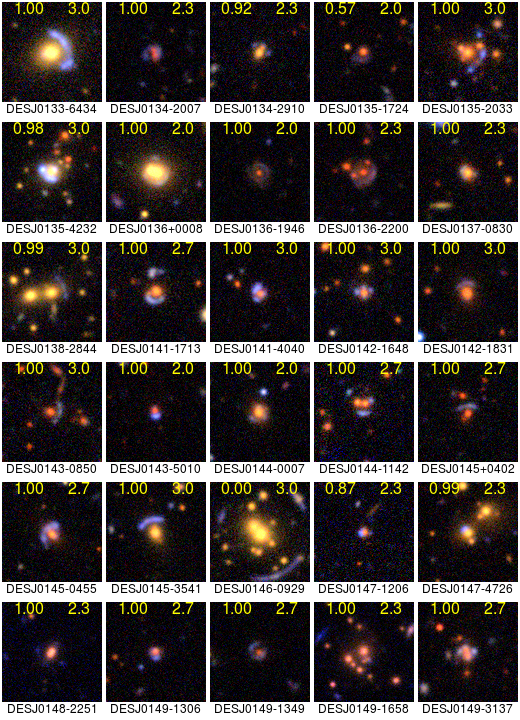}
\caption{Candidate lenses found in DES using
CNNs. In yellow, left: best CNN score, right: human grade.}\label{fig:candidate-lenses-3}
\end{figure*}
\begin{figure*}
\centering
\includegraphics[width=0.80000\textwidth]{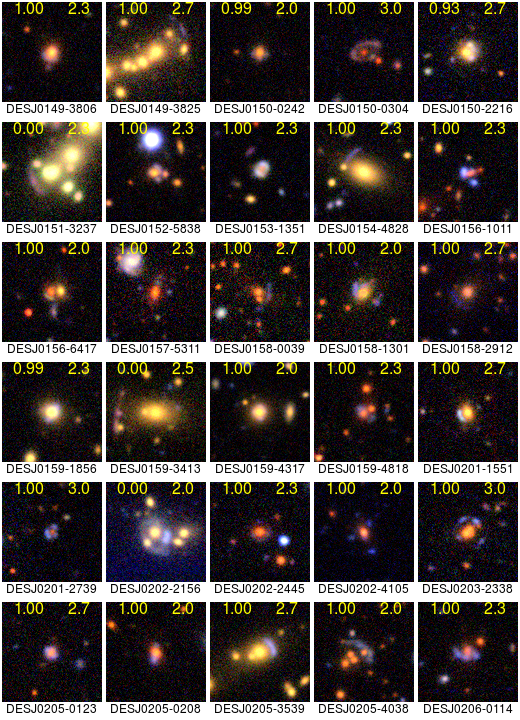}
\caption{Candidate lenses found in DES using
CNNs. In yellow, left: best CNN score, right: human grade.}\label{fig:candidate-lenses-4}
\end{figure*}
\begin{figure*}
\centering
\includegraphics[width=0.80000\textwidth]{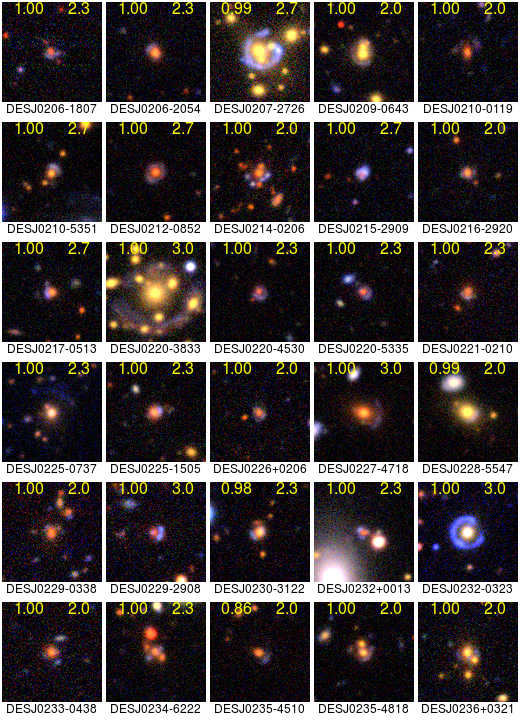}
\caption{Candidate lenses found in DES using
CNNs. In yellow, left: best CNN score, right: human grade.}\label{fig:candidate-lenses-5}
\end{figure*}
\begin{figure*}
\centering
\includegraphics[width=0.80000\textwidth]{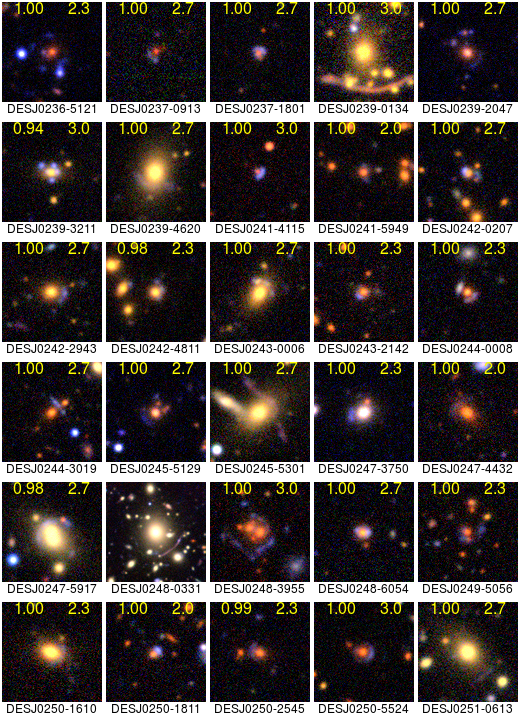}
\caption{Candidate lenses found in DES using
CNNs. In yellow, left: best CNN score, right: human grade.}\label{fig:candidate-lenses-6}
\end{figure*}
\begin{figure*}
\centering
\includegraphics[width=0.80000\textwidth]{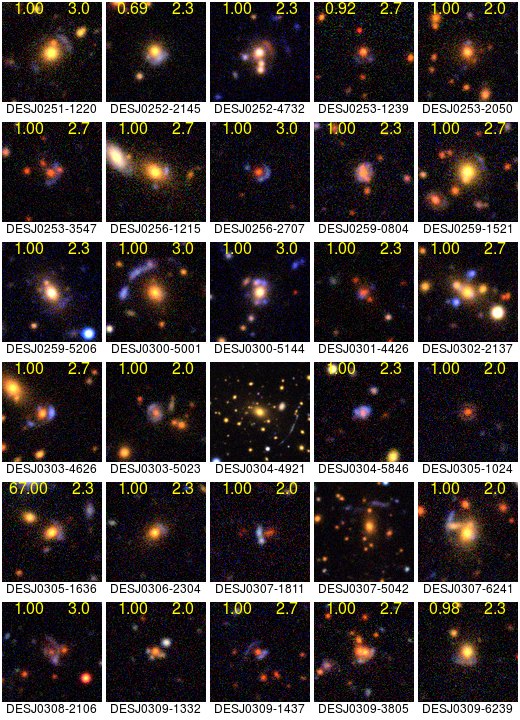}
\caption{Candidate lenses found in DES using
CNNs. In yellow, left: best CNN score, right: human grade.}\label{fig:candidate-lenses-7}
\end{figure*}
\begin{figure*}
\centering
\includegraphics[width=0.80000\textwidth]{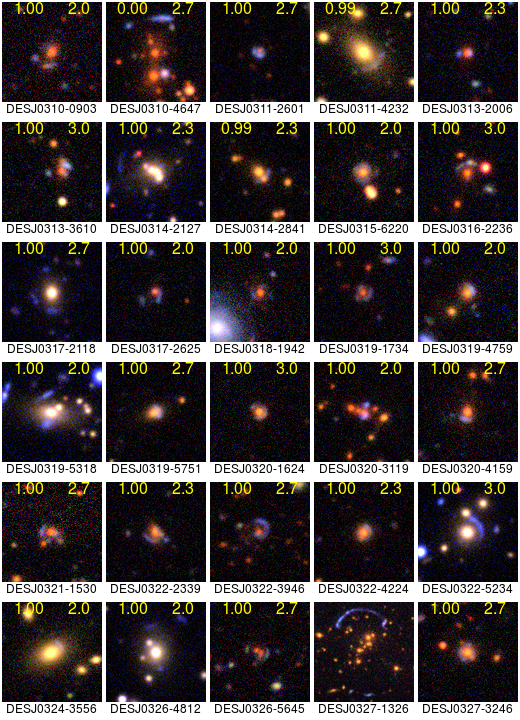}
\caption{Candidate lenses found in DES using
CNNs. In yellow, left: best CNN score, right: human grade.}\label{fig:candidate-lenses-8}
\end{figure*}
\begin{figure*}
\centering
\includegraphics[width=0.80000\textwidth]{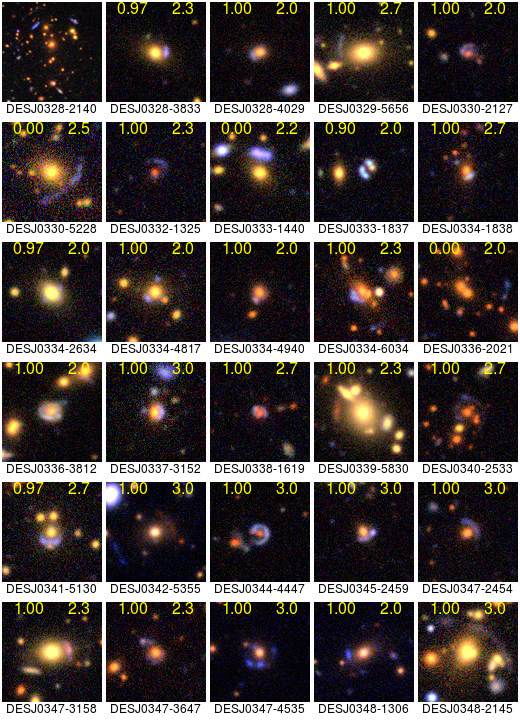}
\caption{Candidate lenses found in DES using
CNNs. In yellow, left: best CNN score, right: human grade.}\label{fig:candidate-lenses-9}
\end{figure*}
\begin{figure*}
\centering
\includegraphics[width=0.80000\textwidth]{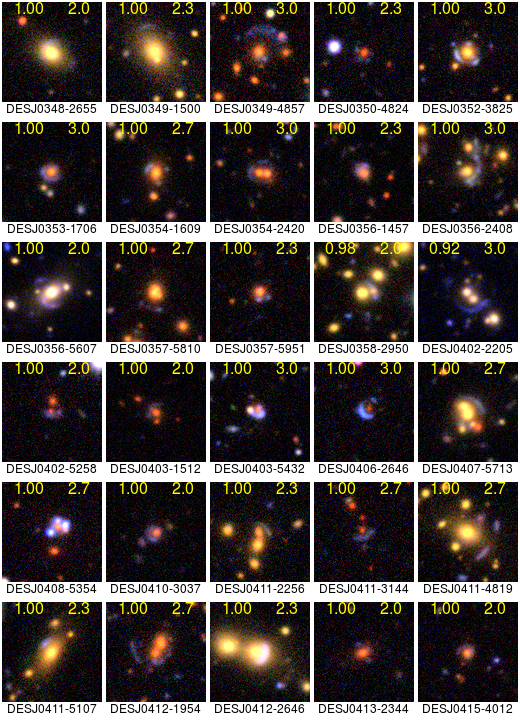}
\caption{Candidate lenses found in DES using
CNNs. In yellow, left: best CNN score, right: human grade.}\label{fig:candidate-lenses-10}
\end{figure*}
\begin{figure*}
\centering
\includegraphics[width=0.80000\textwidth]{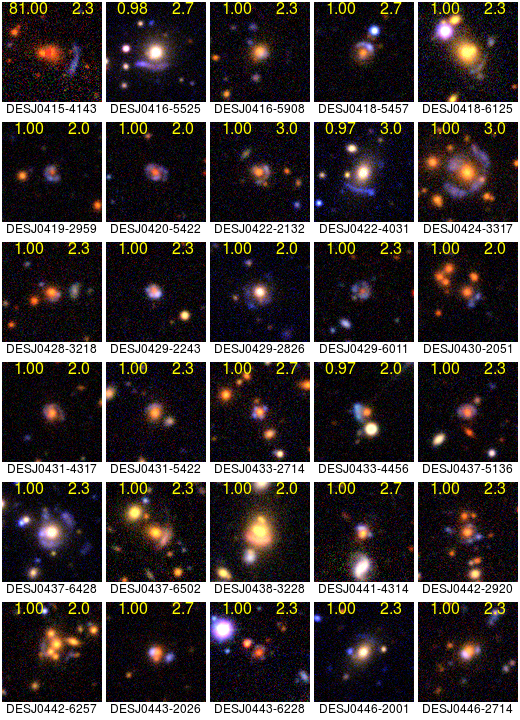}
\caption{Candidate lenses found in DES using
CNNs. In yellow, left: best CNN score, right: human grade.}\label{fig:candidate-lenses-11}
\end{figure*}
\begin{figure*}
\centering
\includegraphics[width=0.80000\textwidth]{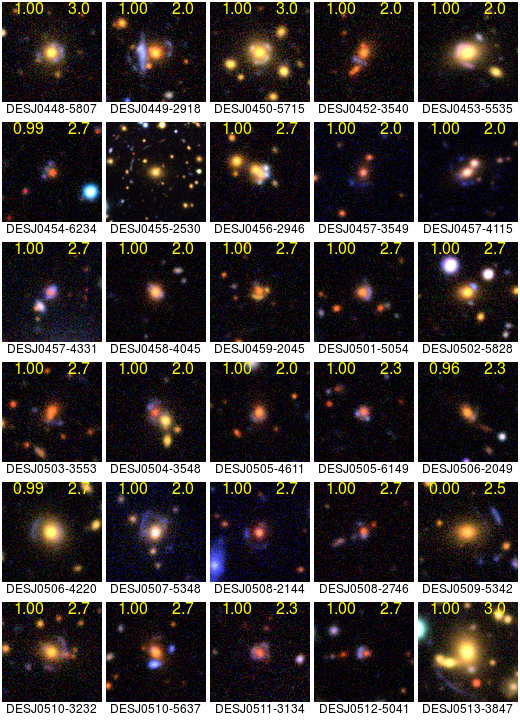}
\caption{Candidate lenses found in DES using
CNNs. In yellow, left: best CNN score, right: human grade.}\label{fig:candidate-lenses-12}
\end{figure*}
\begin{figure*}
\centering
\includegraphics[width=0.80000\textwidth]{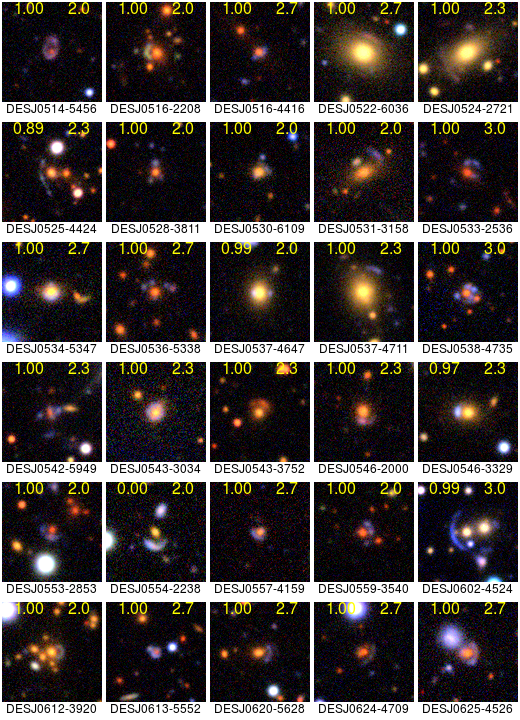}
\caption{Candidate lenses found in DES using
CNNs. In yellow, left: best CNN score, right: human grade.}\label{fig:candidate-lenses-13}
\end{figure*}
\begin{figure*}
\centering
\includegraphics[width=0.80000\textwidth]{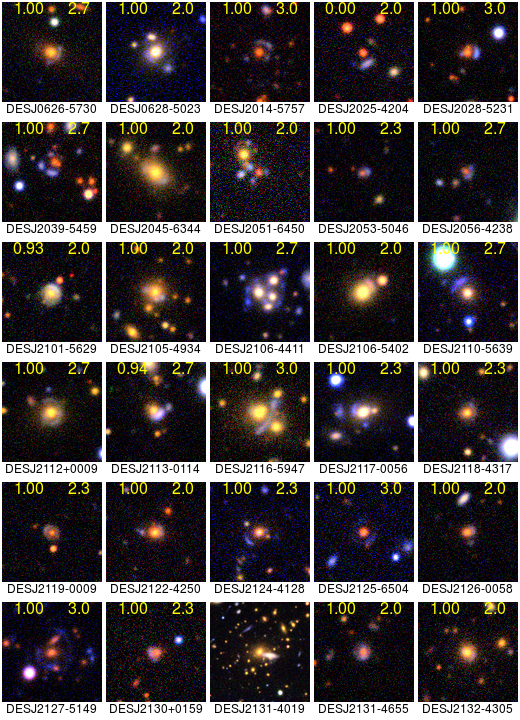}
\caption{Candidate lenses found in DES using
CNNs. In yellow, left: best CNN score, right: human grade.}\label{fig:candidate-lenses-14}
\end{figure*}
\begin{figure*}
\centering
\includegraphics[width=0.80000\textwidth]{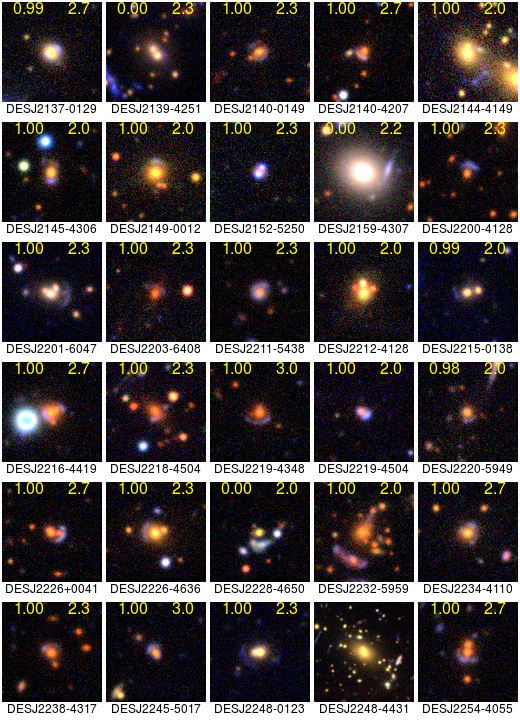}
\caption{Candidate lenses found in DES using
CNNs. In yellow, left: best CNN score, right: human grade.}\label{fig:candidate-lenses-15}
\end{figure*}
\begin{figure*}
\centering
\includegraphics[width=0.80000\textwidth]{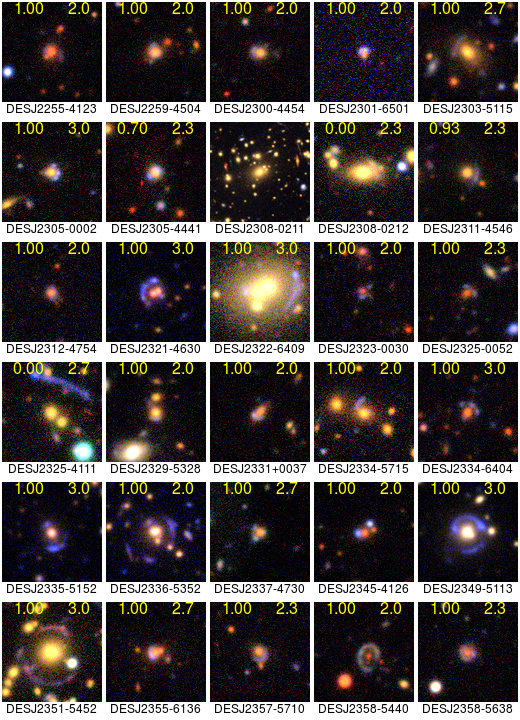}
\caption{Candidate lenses found in DES using
CNNs. In yellow, left: best CNN score, right: human grade.}\label{fig:candidate-lenses-16}
\end{figure*}
\begin{figure*}
\centering
\includegraphics[width=0.80000\textwidth]{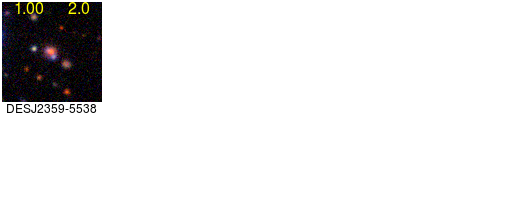}
\caption{Candidate lenses found in DES using
CNNs. In yellow, left: best CNN score, right: human grade.}\label{fig:candidate-lenses-17}
\end{figure*}

\clearpage
\newpage

\appendix
\section{Keras model summary}\label{sec:appendix-1}

\begin{deluxetable}{lll}[h]



\tablecaption{Output of the Keras model summary for the convolutional neural networks used in this lens search}

\tablehead{\colhead{Layer (type)} & \colhead{Output Shape} & \colhead{Param count} \\ 
\colhead{} & \colhead{} & \colhead{} } 

\startdata
conv2d\_13 (Conv2D)  &  (None, 96, 50, 50)  &  34944 \\ \\
max\_pooling2d\_10 (MaxPooling)  &  (None, 96, 24, 24)  &  0 \\ \\
conv2d\_14 (Conv2D)  &  (None, 128, 24, 24)  &  307328 \\ \\
activation\_19 (Activation)  &  (None, 128, 24, 24)  &  0 \\ \\
\enddata
\end{deluxetable}

\clearpage
\newpage

\section{Possible strong lens systems}\label{sec:appendix-2}
\startlongtable
\begin{deluxetable}{llll}
\tablecaption{Possible strong lens systems selected by CNN but with grades 
\(0 < \textrm{grade} < 2\) indicating possible but not probable 
or definite lensing. The object ID is the identifier from the DES Y3A1 coadd object
tables. Where the source has previously been flagged as a potential lens, this is noted. {\footnotesize 
{\sc REFERENCES.} $^1$~\citet{diehlBrightArcsSurvey2017}, $^2$~\citet{petrilloLinKSDiscoveringGalaxyscale2019},
$^3$~\citet{sonnenfeldSuGOHI2018},
$^4$~\citet{wongSurveyGravitationallyLensed2018},
$^5$~\citet{sonnenfeldSL2SGalaxyscaleLens2013}
}}
\tablehead{
    \colhead{Object ID} & 
    \colhead{RA} & 
    \colhead{dec} &
    \colhead{notes}
}
\startdata
140003287 & 00 02 29.66 & -52 29 19.90 &   \\
141268015 & 00 02 41.13 & 02 48 52.85 &   \\
142078524 & 00 03 38.47 & -51 55 57.40 &   \\
142345819 & 00 04 28.14 & -38 44 08.95 &   \\
182976851 & 00 06 10.34 & -55 07 50.09 &   \\
182217410 & 00 06 25.41 & -54 24 33.77 &   \\
182434686 & 00 08 35.18 & -39 22 38.50 &   \\
182361791 & 00 09 47.88 & -05 08 13.06 &   \\
\enddata

\end{deluxetable}



\end{document}